\documentclass[final]{llncs}

\usepackage{etex}
\usepackage{changebar}

\usepackage{amsmath}
\usepackage{stmaryrd}
\usepackage{hyperref}
\usepackage{graphicx}
\usepackage{subcaption} % for subfigures
\usepackage[inference]{semantic}
\usepackage{color}
\usepackage{amssymb}
\usepackage{todonotes}
\usepackage{wasysym}
\usepackage{listings}
\usepackage{pgfplots}
\usepackage{textcomp}
\usepackage{booktabs}
\usepackage{multirow}
\usepackage{rotating}
\usepackage[normalem]{ulem}
\newcommand{\mi}[1]{\mathit{#1}}
\newcommand{\surbracket}[1]{\llbracket #1 \rrbracket}

\newcommand{\GVar}{{\it IVar}}

\newcommand{\LVar}{{\it LVar}}

\newcommand{\Nat}{{\it Int}}
\newcommand{\counter}{{\it count}}

\newcommand{\todec}[2]{\to^{#1}_{#2}}

\newcommand{\procq}{\mathcal{Q}}
\newcommand{\toH}{\rightarrowtail}

\newcommand{\nextOsymb}{\mi{nextObject}}
\newcommand{\nextO}[2]{\nextOsymb({#1},{#2})}
\newcommand{\newQadd}{\mi{newQ_{add}}}

\newcommand{\newQdel}{\mi{newQ_{del}}}

\newcommand{\costmodel}{\mathcal{M}}
\newcommand{\trace}{\mathcal{T}}
\newcommand{\cost}{\mathcal{C}}

\newcommand{\code}[1]{\lstinline[mathescape,basicstyle=\ttfamily]!#1!}
\newcommand{\updL}{\mi{updL}}
\newcommand{\secbeg}{\vspace{-0.1cm}}
\newcommand{\trStmt}[2]{{}^s\surbracket{#1}_{#2}}
\newcommand{\trB}[1]{{}^B\surbracket{#1}}
\newcommand{\trV}[1]{{}^V\surbracket{#1}}

\newcommand{\trMethod}[1]{{}^m\surbracket{#1}}
\newcommand{\trConf}[1]{{}^c\surbracket{#1}}
\newcommand{\trQueue}[1]{{}^q\surbracket{#1}}
\newcommand{\trConfInv}[1]{{}^c\surbracket{#1}^{-1}}
\newcommand{\trQueueInv}[1]{{}^q\surbracket{#1}^{-1}}
\newcommand{\trStmtInv}[1]{{}^s\surbracket{#1}^{-1}}

\newcommand{\getVal}[2]{\mathtt{getVal}({#1},{#2})}

\lstdefinelanguage{hs}
{morekeywords={data,type},
 morecomment=[l]{--},
 moredelim=[is][\uwave]{_}{_},
 moredelim=[is][\uuline]{+}{+}
}

\lstdefinelanguage{abs}
{morekeywords={}, %To avoid differences between the syntax and the program with keywords
 otherkeywords={!},
 morecomment=[l]{//}
}

\begin{document}
	
\nochangebars %Disable change bars

\title{A Formal, Resource Consumption-Preserving Translation of Actors to 
Haskell\thanks{This work was funded partially by the EU project FP7-ICT-610582
 ENVISAGE: Engineering Virtualized Services
 (\url{http://www.envisage-project.eu}), by the Spanish MINECO projects
 TIN2012-38137 and TIN2015-69175-C4-2-R, and by the CM project
 S2013/ICE-3006.}}

\author{Elvira Albert\inst{1} \and Nikolaos Bezirgiannis\inst{2} \and
  Frank de Boer\inst{2} \and
\\ Enrique Martin-Martin\inst{1}}

\institute{
Universidad Complutense de Madrid, Spain\\
\texttt{elvira@sip.ucm.es}, \texttt{emartinm@ucm.es}
\and
Centrum Wiskunde \& Informatica (CWI), Amsterdam, Netherlands\\
\{\texttt{n.bezirgiannis}, \texttt{f.s.de.boer}\}\texttt{@cwi.nl}
}

\maketitle

\begin{abstract}
  We present 
  %and discuss 
  a formal translation of an
  actor-based language with \emph{cooperative scheduling} to the functional language Haskell. The translation is
  proven correct with respect to a formal semantics of the source
  language and a high-level operational semantics of the target,
  i.e. a subset of Haskell.  The main
  correctness theorem is expressed in terms of a simulation relation
  between the operational semantics of actor programs and
  their translation. 
  \cbstart This allows us to then prove that the resource consumption
  is preserved over this translation, as we establish
  an equivalence of the cost of the original and
  Haskell-translated execution traces.  
  \cbend

%discuss how to extend the main result to distributed implementations of the actor-based language.
\end{abstract}

%% Keywords submitted
%% actor model
%% futures
%% cooperative multitasking
%% coroutine
%% continuation
%% functional programming
%% operational semantics

% ;; -*- coding: iso-latin-1; TeX-PDF-mode: t; TeX-master: "0_main" -*-%t
% !TeX root = 0_main.tex

\section{Introduction}
\secbeg\secbeg
% \todo{byE: We need to write a motivation less focused on ABS, and
%   that emphasizes the advantages of translating to Haskell, the
%   difficulties in doing so and the resource-preservation aspect}
% \todo{byN: for the full-ABS compiler the advantage of Haskell is that
% it is closer to the ABS language (functional core and types).
% For our subset is the coroutines (continuations) which are
% a language feature and not a system-level feature, so it can be
% mathem. reasoned}

%% what is actor programming
\emph{Abstract Behavioural Specification} (ABS)~\cite{JohnsenHSSS10} is a formally-defined
language for modeling actor-based programs. 
%We consider actor systems \cite{actors,DBLP:journals/tcs/HallerO09}, a
%model of concurrent programming that has been gaining popularity and
%that is being used in many systems (such as ActorFoundry,
%Asynchronous Agents, Charm++, E, ABS, Erlang, and Scala).  
An actor program consists of computing entities called \emph{actors}, each with a
private state, and thread of control. Actors can communicate by exchanging
messages asynchronously, i.e. without waiting for message delivery/reply.
%
%% what is ABS' concurrent object, how it relates to actors
In ABS, the notion of actor corresponds to the \emph{active object}, 
where objects are the concurrency units, i.e. each object conceptually has a
dedicated thread of execution. Communication is based on asynchronous method calls
where the caller object does not wait for the callee to reply with the method's return value.
Instead, the object can later use a \emph{future} variable \cite{FlanaganF1995,deBoerDJ2007} to extract the result of the asynchronous method.
Each asynchronous method call adds a new \emph{process} to the callee object's process queue.
% The object continuously selects and runs a process from its process queue,
% with the restriction that, at any point in time, only 1 process may be running inside an object (the active process).
ABS supports \emph{cooperative scheduling}, which means that
inside an object, the active process can decide to
explicitly suspend its execution so as to allow another process from
the queue to execute. This way, the interleaving of processes inside an
active object is textually controlled by the programmer, similar to \cbstart coroutines~\cite{Knuth73_book}. \cbend However,
flexible and state-dependent interleaving is still supported: in particular, a process may suspend its execution waiting for a reply to a
method call. 

Whereas ABS has successfully been used to model~\cite{wong_abs_2012}, analyze~\cite{SACO}, and verify~\cite{JohnsenHSSS10} actor programs,
the ``real'' execution of such programs has been a struggle, attributed
to the fact that implementing cooperative scheduling efficiently can be hard
(common languages as Java and C++ have to resort to instrumentation techniques, e.g. fibers~\cite{srinivasan2008kilim}).
This led to the creation of numerous ABS backends with
different cooperative scheduling implementations:\footnote{See \url{http://abs-models.org/documentation/manual/\#-abs-backends} for more information about ABS backends.}
ABS$\rightarrow$Maude using an interpreter and term rewriting,
ABS$\rightarrow$Java using heavyweight threads and manual stack management,
ABS$\rightarrow$Erlang using lightweight threads and thread parking, ABS$\rightarrow$Haskell using lightweight threads and continuations.

%Implementing cooperative scheduling can be non-trivial, even for
%modern high-level programming languages (e.g. Java, C++)
%because of their  stack-based nature.
%A recent relevant technology  is to use \emph{fibers} \cite{srinivasan2008kilim}, which adds 
%support for cooperative threads by instrumenting low-level code (commonly via bytecode manipulation)
%to save and restore parts of the stack.
%We instead opted for source-to-source translating ABS programs to Haskell, a functional language with language-level support for coroutines, 
%based on the hypothesis that a high-level translation
%serves as a better middleground between execution speed and most importantly semantic correctness.
%Our transcompiler translates ABS programs % written in the familiar direct style
%to equivalent Haskell-code, % in \emph{continuation-passing style} (CPS)
%which is then compiled to native code by a Haskell compiler and executed. 
%Prior alternative approaches for executing ABS have been an Erlang translator, that utilizes Erlang's preemptive lightweight processes
%to simulate cooperative threads, and a Java translator, that manages a global dynamic pool of heavyweight threads. 
%\todo{Mention Maude backend?}

The overall contribution of this paper is a formal,
resource-consumption preserving translation of the
concurrency subset of the ABS language into Haskell, given
as an adaptation of the canonical ABS$\rightarrow$Haskell backend~\cite{bezirg_cloud_2016}. 
We opted for the Haskell backend relying on the hypothesis that Haskell 
serves as a better middleground between execution speed and most importantly semantic correctness. The translation is based on compiling ABS methods into Haskell functions with
\emph{continuations}---similar transformations have been performed in the
actor-based Erlang language wrt. rewriting systems~\cite{PalaciosV15,Vidal13} and rewriting logic~\cite{Noll01}, 
in the translation of ABS to Prolog~\cite{AlbertAGZ12} 
and a subset of ABS  to Scala~\cite{nakata_compiling_2013}.  
%We take a modified extract of the canonical ABS-to-Haskell compiler focusing on cooperative scheduling;
However, what is unique in our translation and constitutes our
main contribution, is that the translation is resource preserving as we prove
in two steps:
\begin{itemize} 
\item{\em Soundness.} We provide a formal statement of the soundness
  of this translation of ABS into Haskell which is expressed in terms
  of a simulation relation between the operational ABS semantics and
  the semantics of the generated Haskell code. The soundness claim
  ensures that every Haskell derivation has an equivalent one in
  ABS. However, since for efficiency reasons, the translation
  fixes a selection order between the objects and the processes within
  each object, we do not have a completeness result.
\item{\em Resource-preservation.} 
  As a corollary we have that the 
  % We prove formally that the
  transformation preserves the resource consumption, i.e., 
  %given a \emph{cost model} that assigns a cost to each ABS instruction, 
  %we prove that 
  the cost of the Haskell-translated
  program is the same as the original ABS program wrt. any 
  \emph{cost model} that assigns a cost to each ABS instruction,
  since both programs execute the same trace of ABS instructions. This result allows us to
  ensure that upper bounds on the resource consumption obtained by the
  analysis of the original ABS program are preserved during
  compilation and are thus valid bounds for the
  Haskell-translated program as well.
\end{itemize}
In Section~\ref{sec:lang} we specify the syntax of the source language and detail
its operational semantics. Section~\ref{sec:targetlang} describes our target language and defines
the compilation process. We present the correctness and resource preservation 
results in Section~\ref{sec:correctness}, as well as the intermediate semantics used in this process.
In Section~\ref{sec:experiments} we show that the runtime environment does not
introduce any significant overhead when executing ABS instructions, and show that
the upper bounds obtained by the cost analysis are sound. Finally, Section~\ref{sect:conclusion} 
contains the conclusions and future work. Complete proofs of the theoretical results can be found at \url{http://gpd.sip.ucm.es/enrique/publications/lopstr16_ext.pdf}.
\section{Source language}\label{sec:lang}
\secbeg\secbeg
\begin{figure}[tbp]
\begin{minipage}{0.45\textwidth}
%\hspace*{1cm}
%\begin{minipage}{5cm}
%\begin{align*}
$\begin{array}{ll}
S ::=&\ x \texttt{:=} E ~|~ f \texttt{:=} x \texttt{!} m( \bar{y} ) \\
     & |~ \mathtt{await}\ f ~|~ \mathtt{skip} ~|~ \mathtt{return}\ z \\
     & |~ S_1 \mathtt{;} S_2 ~|~ \mathtt{if}\ B\ \{ S \} \ \mathtt{else}\  \{ S \} \\
     & |~ \mathtt{while} \ B\ \{ S \} \\
%\end{align*}
%\end{minipage}
%\begin{minipage}{5cm}
% \begin{align*}
%E ::=&\ x ~|~ r ~|~ \mathtt{new} ~|~ f\mathtt{.get} ~|~ m( \bar{y} ) \\
E ::=&\ V ~|~ \mathtt{new} ~|~ f\mathtt{.get} ~|~ m( \bar{y} ) \\
V ::=&\ x ~|~ r ~|~ I \\
 B ::=&\ B \land B ~|~ B \lor B ~|~ \lnot B ~|~  V \equiv V  \\ 
D ::=&\ m ( \bar{r} )\{\;S\;\}\\
P ::=&\ \overline{D} ~:~ \mathtt{main}()\{\;S\;\}
%\end{align*}
\end{array}$
%\end{minipage}
%
\end{minipage}
\begin{minipage}{0.3\textwidth}
\begin{lstlisting}[language=abs,basicstyle=\small\ttfamily]
main() {
  node1 = new; *'\label{ex:new1}'*
  node2 = new; *'\label{ex:new2}'*
  f1 = node1!map(v*'$_1$'*); *'\label{ex:async1}'*
  f2 = node2!map(v*'$_2$'*); *'\label{ex:async2}'*
  await f1; *'\label{ex:await1}'*
  await f2; *'\label{ex:await2}'*
  r1 = f1.get; *'\label{ex:get1}'*
\end{lstlisting}
\end{minipage}
\begin{minipage}{0.2\textwidth}
\begin{lstlisting}[language=abs,firstnumber=9,basicstyle=\small\ttfamily]
 r2 = f2.get; *'\label{ex:get2}'*
 r = reduce(r1,r2); *'\label{ex:sync}'*
 return r; *'\label{ex:ret}'*}

map(v) { 
  ... }
reduce(v1,v2) { 
  ... }
\end{lstlisting}
\end{minipage}
\secbeg\secbeg
\caption{(a) syntax of source language (b) a simplified MapReduce task in ABS}
\label{fig:src}
\end{figure}

%\todo{EM: Figure~\ref{fig:running} can be reduced removing one task and one object}
%\todo{The text in this section should be reduced and an example added}

Our language  is based on ABS~\cite{JohnsenHSSS10}, 
a statically-typed, actor-based language
with a purely-functional core (ADTs, functions, parametric polymorphism)
and an object-based imperative layer: objects with private-only attributes,
and interfaces %(sporting inheritance) 
that serve as types to the objects.
ABS extends the OO paradigm with support for \emph{asynchronous} method calls;
each call results in a new \emph{future} (placeholder for the method's result) returned to the caller-object, and a new process
(stored in the callee-object's process queue) which runs the method's activation.
The active process inside an object (only one at any given time) may
decide to explicitly suspend its execution so as to allow another process from the same queue to execute.

In this paper, we simplify ABS to its subset that concerns the concurrent interaction of processes (inside and between objects), so as to focus 
solely on the more challenging part of proving correctness of the cooperative concurrency.
In other words, the ABS language is stripped of its functional core, local variables, object groups~\cite{schafer2010jcobox} and types (we assume the input programs are well-typed w.r.t ABS type-system).
%In this paper, we deal with a subset of ABS concerning the concurrent interaction
%of processes both inside and between objects; in a nutshell, the language is stripped of its functional core and all types/interfaces.
%The choice here was made to focus on the more challenging part of proving correctness of the cooperative concurrency.
% However, the full-blown compiler that we will use in the experiments
% deals with the whole ABS language.
The formal syntax of the statements $S$ of the subset is shown in Fig.~\ref{fig:src}(a).
Values in our subset are references (object or futures) and integer numbers; values can be stored in method's formal parameters or attributes.
We syntactically distinguish between method parameters $r$ and attributes. The attributes are further distinguished 
for the values they hold: attributes holding object references or integer values (denoted by $x,y,z\ldots$), and future attributes holding future references (denoted by $f$).
An assignment $f$\texttt{:=}$x$\texttt{!}$m(\bar{y})$ stores to the future attribute $f$ a new future reference returned by asynchronously calling the method $m$ on the object attribute $x$
passing as arguments the values of object attributes $\bar{y}$. An assignment $x$\texttt{:=}$E$ stores to an object attribute the result of executing the right-hand side $E$. A right-hand side can be the value of a method parameter $r$, an attribute $x$, an integer expression $I$ (an integer value, addition, subtraction, etc.),
%\footnote{We assume that integer expressions only involve integer values, so they cannot contain futures or object references.}
a reference to a new object \texttt{new}, the result of a synchronous same-object method call
$m(\bar{y})$, or the result of an asynchronous method call  $f$.\code{get}  stored in the future attribute $f$.
A call to  $f$.\code{get}  will block the object and all its processes until the result of the asynchronous call is ready.
The statement $\mathtt{await}~f$ may be used (usually before calling  $f$.\code{get} ) to instead release the current process until the result of $f$ has been computed,
allowing another same-object process to execute. Sequential composition of two statements $S_1$ and $S_2$ is denoted by $S_1;S_2$.
The Boolean condition $B$ in the \emph{if} and \emph{while} statement is a Boolean combination of reference equality between values of attributes.
Again, note that, we assume expressions to be well-typed: integer expressions
cannot contain futures or object references and boolean equality is between
same-type values.
The statement $\mathtt{return}\; z$ returns the value of the attribute $z$ both in synchronous and asynchronous method calls.
A method declaration $D$ maps a method's name and formal parameters to a statement $S$ (method body).
We consider that every method has one \texttt{return} and it is the final statement.
Finally, a program $P$ is a set of method declarations $\bar{D}$ and a special method \texttt{main} that has no formal
parameters and acts as the program's entry point. 

The program of Fig.~\ref{fig:src}(b) shows a basic version of a MapReduce task~\cite{Dean2008}
implemented using actors in ABS. For clarity the example uses only two \emph{map} nodes and 
a single \emph{reduce} computation performed in the controller node (the actor
running \code{main}). First the controller creates two objects \code{node1} and \code{node2} (L\ref{ex:new1}--L\ref{ex:new2}), and invokes asynchronously \texttt{map} with different values \code{v}$_1$ and \code{v}$_2$ (L\ref{ex:async1}--L\ref{ex:async2}). In MapReduce,
all \texttt{map} invocations must finish before executing the \emph{reduce} phase: therefore, the \code{await} instructions in 
L\ref{ex:await1}--L\ref{ex:await2} wait for the termination of the two calls to \texttt{map}, releasing the processor so that any other process 
in the same object of \code{main} can execute. Once they have finished, the \code{get} statements in L\ref{ex:get1}-L\ref{ex:get2} obtain the results from the futures \code{f1} and \code{f2}. Although \code{get} statements block the object (in this case \emph{main}) and all of its processes until the result is ready, this does not occur in our example because the preceding \code{await}s assure the result is available. Finally, L\ref{ex:sync} contains a synchronous-method self call to \code{reduce} that combines the partial results from the \emph{map} phase.

\subsection{Operational semantics}\label{sec:operationalSem}
\secbeg
\begin{figure}[tb]
{\footnotesize
$$
\inference[(\textsc{Assign})]
{
%\mathtt{getVal}(h(n),V) & h' = h[ (n)(x) \mapsto h(n)(y)] 
\getVal{h(n)}{V} = v & h' = h[ (n)(x) \mapsto v)] 
}
{
\langle n: (\texttt{x:=}V;S,l) \cdot Q, h \rangle \rightarrow \langle n: (S,l) \cdot Q, h' \rangle
}
$$
%\subsubsection{Local rules}
% $$
% \inference[(\textsc{Assign I})]
% {
% h' = h[ (n)(x) \mapsto h(n)(y)] 
% }
% {
% \langle n: (\texttt{x:=y};S,l) \cdot Q, h \rangle \rightarrow \langle n: (S,l) \cdot Q, h' \rangle
% }
% $$
% 
% $$
% \inference[(\textsc{Assign II})]
% {
% h' = h[ (n)(x) \mapsto r] 
% }
% {
% \langle n: (\texttt{x:=r};S,l) \cdot Q, h \rangle \rightarrow \langle n: (S,l) \cdot Q, h' \rangle
% }
% $$

$$
\inference[(\textsc{New})]
{
h(\mi{count}) = m & h' = h[ (n)(x) \mapsto m, (m) \mapsto \epsilon, \mi{count} \mapsto m + 1] 
}
{
\langle n: (\texttt{x:=new};S,l) \cdot Q, h \rangle \rightarrow \langle n: (S,l) \cdot Q, h' \rangle
}
$$

$$
\inference[(\textsc{Get})]
{
h(h(n)(f)) \neq \bot & h' = h[(n)(x) \mapsto h(h(n)(f))] \\
}
{
\langle n: (\texttt{x:=f.get};S,l) \cdot Q, h \rangle \rightarrow \langle n: (S,l) \cdot Q, h' \rangle
}
$$

$$
\inference[(\textsc{Await I})]
{
h(h(n)(f)) \neq \bot \\
}
{
\langle n: (\texttt{await f};S,l) \cdot Q, h \rangle \rightarrow \langle n: (S,l) \cdot Q, h \rangle
}
$$

$$
\inference[(\textsc{Await II})]
{
h(h(n)(f)) = \bot \\
}
{
\langle n: (\texttt{await f};S,l) \cdot Q, h \rangle \rightarrow \langle n: Q \cdot (\texttt{await f};S,l), h \rangle
}
$$

$$
\inference[(\textsc{Async})]
{
h(n)(x) = d & h(\mi{count}) = l' & \bar{v} = h(n)(\bar{z}) \\
h' = h[(n)(f) \mapsto l', (l') \mapsto \bot,\mi{count} \mapsto l'+1]
}
{
\langle n: (\texttt{f:=x!m($\bar{z}$)};S,l) \cdot Q, h \rangle \stackrel{d.m(l',\bar{v})}{\longrightarrow} \langle n: (S,l) \cdot Q, h' \rangle
}
$$

$$
\inference[(\textsc{Sync})]
{
(m(\bar{w}) \mapsto S_m) \in D & \tau = [\bar{w} \mapsto h(n)(\bar{z})] & S' = \widehat{(S_m\tau)}^x
% (m(\bar{w}) \mapsto S_m) \in D~\mathit{fresh} & \tau = [\bar{w} \mapsto eval(\bar{z},n,h)] & S' = \widehat{(S_m\tau)}^x
}
{
\langle n: (\texttt{x:=m($\bar{z}$)};S,l) \cdot Q, h \rangle \rightarrow \langle n: (S';S,l) \cdot Q, h \rangle
}
$$

$$
\inference[(\textsc{Return$_A$})]
{
h' = h[(l) \mapsto h(n)(x)]
}
{
\langle n: (\texttt{return}^* \texttt{x};S,l) \cdot Q, h \rangle \rightarrow \langle n: Q, h' \rangle
}
$$

$$
\inference[(\textsc{Return$_S$})]
{
h' = h[ (n)(z) \mapsto h(n)(x)] 
}
{
\langle n: (\texttt{return}^z~\texttt{x};S,l) \cdot Q, h \rangle \rightarrow \langle n: (S,l) \cdot Q, h' \rangle
}
$$
}
%\rule{5cm}{1pt}
\secbeg\secbeg\secbeg
\caption{Operational semantics: Local rules\label{fig:localSem}}
\end{figure}

In order  to describe the operational semantics of the language defined above
we first introduce the following concepts and assumptions.
% We use the set $\Nat$ of  integer numbers  to represent integer values and also to 
% encode object and future references,
% used to identify dynamically  generated objects and futures.
The values considered in this paper are in the $\Nat$ set: integer constants and 
 dynamically generated references to objects and futures.
%By $\ObjectRefs$ we denote an infinite set of object  references used to identify objects and by $\FutRefs$ we denote an infinite set  references for the identification of futures.
% We denote by $\Sigma=\GVar\rightarrow \Nat$ the set of assignments of references,
% presented by natural numbers,
% to  the instance variables (of an object), with typical element $\sigma$ and empty element $\epsilon$.
We denote by $\Sigma=\GVar\rightarrow \Nat$ the set of assignments of values
to  the instance variables (of an object), with typical element $\sigma$ and empty element $\epsilon$.
A closure consists of a statement $S$ obtained by replacing
its free variables by actual values (note that variables are introduced as method parameters and can only appear in $E$)
and a future reference, represented by an integer, for storing the return value.
By $S\tau$, where $\tau\in \LVar\rightarrow \Nat$, we denote the instantiation 
obtained from $S$ by replacing each variable $x$ in $S$ by $\tau(x)$. 
Finally, we represent the global heap $h$ by a triple  $(n,h_1,h_2)$
consisting of a natural number $n$ and
\emph{partial} functions (with finite disjoint domains)
$h_1:\Nat\rightarrow\Sigma$ and $h_2:\Nat \rightarrow \Nat_\bot$, 
where $\Nat_\bot=\Nat\cup\{\bot\}$ ($\bot$ is used to denote ``undefined'').
The number $n$ is used to generate references to new objects and futures.
The function $h_1$ specifies for each existing object, i.e., a number $n$ such $h_1(n)$ is defined,
its \emph{local} state.
The function $h_2$ specifies for each existing future reference, i.e.,
a number $n$ such $h_2(n)$ is defined,
its return value (absence of which is indicated by $\bot$).
In the sequel we will simply denote the first
component of $h$ by $h(\counter)$, and write $h(n)(x)$, instead of $h_1(n)(x)$,
and $h(n)$, instead of $h_2(n)$. 
We will use the notation $h[\mi{count} \mapsto n]$ to generate a heap equal to $h$ but with the counter set to $n$. A similar notation $h[n \mapsto \bot]$ will be used for future variables, $h[(n)(x) \mapsto v]$ for storing the value $v$ in the variable $x$ in object $n$ and $h[n \mapsto \epsilon]$ for initializing the mapping of an object.

An object's \emph{local} configuration denoted by the (object) reference $n$  consists
of a pair  $\langle n: Q,h\rangle$ where $Q$ is a list of closures and $h$ is the global heap. 
We use $\cdot$ to concatenate lists, i.e., $(S,l)\cdot Q$ represents a list where $(S,l)$ is the head and $Q$ is the tail.
A \emph{global} configuration---denoted with the letters $A$ and $B$---is a pair $\langle C,h\rangle$ containing a set of lists of closures $C=\{\overline{Q}\}$ and a global heap $h$. Fig.~\ref{fig:localSem} contains the relation that describes the local behavior of an object (omitting the standard rules
for sequential composition, \code{if} and \code{while} statements). Note that the first closure of the list $Q$ is the active process of the object, so the different rules process the first statement of this closure. When the active process finishes or releases the object in an \texttt{await} statement, the next process in the list will become active, following a FIFO policy. 
% The rules \textsc{(Assign I)} and \textsc{(Assign II)} modify the heap storing the new value of variable $x$ of object $n$.
The rule \textsc{(Assign)} modifies the heap storing the new value of variable $x$ of object $n$. It uses the function $\getVal{\Sigma}{V}$ to evaluate
an expression $V$ involving integer constants and variables in the object's current state $\Sigma$.
The \textsc{(New)} rule stores a new object reference in variable $x$, increments the counter of objects references and inserts an empty mapping $\epsilon$ for the variables of the new object $m$. Rule \textsc{(Get)} can only be applied if the future is available, i.e., if its value is not $\bot$. In that case, the value of the future is stored in the variable $x$. Both rules \textsc{(Await I)} and \textsc{(Await II)} deal with \texttt{await} statements. If the future $f$ is available, it continues with the same process. Otherwise it moves the current process to the end of the queue,  thus avoiding starvation. Note that the \texttt{await} statement is not consumed, as it must be checked when the process becomes active again.
When invoking the method $m$ asynchronously in rule \textsc{(Async)} the destination object $d$ and the values of the parameters $\bar{r}$ are computed. Then a new future reference $l$ initialized to $\bot$ is stored in the variable $f$, and the counter is incremented. The information about the new process that must be created is included as the decoration $d.m(l',\bar{v})$  of the step. Synchronous calls---rule \textsc{(Sync)}---extend the active task with the statements of the method body, where the parameters have been replaced by their value using the substitution $\tau$. In order to return the value of the method and store it in the variable $x$, the \texttt{return} statement of the body is marked with the destination variable $x$, called \emph{write-back variable}.
%replaced by an assignment of $x$. 
This marking is formalized in the $\hat{\cdot}^s$ function, defined as follows (recall that \texttt{return} is the last
statement of any method):

\secbeg\secbeg
{\footnotesize
\begin{displaymath}
%\begin{equation*}
\widehat{S}^s = \left\{
\begin{array}{rl}
S_1;\widehat{S_2}^s & \text{~~~if } S=S_1;S_2,\\
%\texttt{x}\eqmark\texttt{z} & \text{~~~if } S=\texttt{return z},\\
\texttt{return}^s~\texttt{z} & \text{~~~if } S=\texttt{return z},\\
S & \text{~~~i.o.c}.
\end{array} \right.
%\end{equation*}
\end{displaymath}
}
%We mark the assignment with a $\star$ to syntactically distinguish between original assignments and those generated by returns in method calls.
Rule \textsc{(Return$_A$)} finishes an asynchronous method invocation (in this case the \texttt{return} keyword is marked with *, see rule \textsc{(Message)} in Fig.~\ref{fig:globalSem}), so it removes the current process and stores the final value in the future $l$. On the other hand, rule \textsc{(Return$_S$)} finishes a synchronous method invocation (marked with the write-back variable), so it behaves like a \texttt{z:=x} statement.

\begin{figure}[tb]
$$
\inference[(\textsc{Internal})]
{
\langle n: Q, h \rangle \to \langle n: Q', h' \rangle
}
{
\langle (n:Q) \cup C, h \rangle \rightarrow \langle (n: Q') \cup C, h' \rangle
}
$$
\vspace*{.1cm}
$$
\inference[(\textsc{Message})]
{
\langle n: Q_n, h \rangle \stackrel{d.m(l',\bar{v})}{\longrightarrow} \langle n: Q', h' \rangle \\ m(\bar{w}) \mapsto S_m \in D & \tau = [\bar{w} \mapsto \bar{v}] & S' = \widehat{(S_m\tau)}^*
}
{
\langle (n:Q_n) \cup (d:Q_d) \cup C, h \rangle \rightarrow \langle (n: Q') \cup (d:Q_d\cdot(S',l')) \cup C, h' \rangle
}
$$

\secbeg\secbeg\secbeg
\caption{Operational semantics: Global rules\label{fig:globalSem}}
\end{figure}

Based on the previous rules, Fig.~\ref{fig:globalSem} shows the relation describing the global behavior of configurations. The \textsc{(Internal)} rule applies any of the rules in Fig.~\ref{fig:localSem}, except \textsc{(Async)}, in any of the objects. 
The \textsc{(Message)} rule applies the rule \textsc{(Async)} in any of the objects. It creates a new closure $(\widehat{S_m\tau}^*,l')$  for the new process invoking the method $m$, and inserts it at the back of the list of the destination object $d$. 
Note the use of~~$\widehat{\cdot}^*$ to mark that the \texttt{return} statement corresponds to an asynchronous invocation.
Note that in both \textsc{(Internal)} and \textsc{(Message)} rules the selection of the object to execute is non-deterministic.
%In some situations 
When needed, we decorate both local and global steps with object reference $n$ and statement $S$ executed, i.e., $\langle n:Q,h\rangle \to^n_S \langle n:Q',h'\rangle$ and $\langle C,h\rangle \to^n_S \langle C',h'\rangle$.

We remark that the operational semantics shown in Fig.~\ref{fig:localSem} and~\ref{fig:globalSem}  is very similar to the foundational ABS semantics  presented in~\cite{JohnsenHSSS10}, considering that every object is a \emph{concurrent object group}. The main difference is the representation of configurations: in~\cite{JohnsenHSSS10} configurations are sets of futures and objects that contain their local stores, whereas in our semantics all the local stores and futures are merged in a global heap. Finally, our operational semantics considers a FIFO policy in the processes of an object, whereas~\cite{JohnsenHSSS10} left the scheduling policy unspecified.

\section{Target language}\label{sec:targetlang}
\secbeg\secbeg
Our ABS subset is translated to Haskell with coroutines. \cbdelete%(the translator itself written in Haskell).
A coroutine is a generalization of a subroutine: besides the usual entry-point/return-point of a procedure
a coroutine can have other entry/exit points, at intermediate locations of the procedure's body. Simply put,
a coroutine does not have to run to completion; the programmer can specify places where
a coroutine can suspend and later resume exactly where it left off.

Coroutines can be implemented natively on top of programming languages that support first-class \emph{continuations} (which
subsequently require support for closures and tail-call optimization). 
A continuation with reference to a program's point of execution, is a datastructure that captures what the remaining of the program does (after the point).
As an example, consider the Haskell program at Fig.~\ref{fig:cps}(a).
The continuation of the call to \lstinline+(even 3)+ at L2 is $\lambda$\lstinline+a+$\rightarrow$\lstinline+print a+, assuming \lstinline+a+ is the result of call to \lstinline+even+ and the continuation is represented as a function.
The continuation of \lstinline+(mod x 2)+ at L1 is the function $\lambda$\lstinline+a+$\rightarrow$\lstinline+print(eq a 0)+ where \lstinline+x+ is bound by the \lstinline+even+ function and \lstinline+a+ is the result of \lstinline+(mod x 2)+. Abstracting
over any program, an expression with type \lstinline+expr::a+
has a continuation \lstinline+k+ with type \lstinline+k::(a+$\rightarrow$\lstinline+r)+ with \lstinline+a+ being the expression's result type and \lstinline+r+ the program's overall result type. To benefit from continuations (and thus coroutines), a program has to be transformed in the so-called \emph{continuation-passing style} (CPS):
a function definition of the program \lstinline+f::args+$\rightarrow$\lstinline+a+ is rewritten to take its current continuation as an extra last argument, as in \lstinline+f'::args+$\rightarrow$\lstinline+(a+$\rightarrow$\lstinline+r)+$\rightarrow$\lstinline+r+. A function call is also rewritten to apply this extra argument
with the actual continuation at point. 

A CPS transformation can be applied to all functions of a program, as in the example of Fig.~\ref{fig:cps}(b),
or (for efficiency reasons) to only the subset that relies on continuation support, e.g. only those functions that need to suspend/resume.
For our case, ABS is translated to Haskell with CPS applied only to statements and methods, but not (sub)expressions.
Continuations have the type \lstinline+k::a+$\rightarrow$\lstinline+Stm+ where \lstinline+Stm+ is a recursive datatype with each one of its constructors being a statement, and the recursive position being the statement's current continuation. \lstinline+Stm+ being the  program's overall result type (\lstinline+Stm+$\equiv$\lstinline+r+),
reveals the fact that the translation of ABS constructs
a Haskell AST-like datatype ``knitted'' with CPS (Fig.~\ref{fig:stm}), which will only later be interpreted at runtime (Sec~\ref{sec:runtime}): capturing the
continuation of an ABS process allows us to save the process' state (e.g. call stack)
and rest of statements as data. For technical convenience, our statements and methods do not directly pass results among each
other but only indirectly through the state (heap); thus, we can reduce our continuation type to 
\lstinline+k::()+$\rightarrow$\lstinline+Stm+ and further to the ``nullary'' function \lstinline+k::Stm+.
Accordingly the CPS type of our methods (functions) 
and statements (constructors) becomes \lstinline+f'::args+$\rightarrow$\lstinline+Stm+$\rightarrow$\lstinline+Stm+.
Worth to mention in Fig.~\ref{fig:stm} is that the body of \code{While} statement and the two branch bodies of \code{If} can be thought of as functions with no \lstinline+args+ written also in CPS (thus type \code{Stm} $\to$ \code{Stm}) to ``tie'' each body's last statement to the continuation \emph{after} executing the control structure. 
% NIKOS: note to self: I am not sure if this qualifies as Higher-Order Abstract Syntax
%This style of Higher-Order Abstract Syntax is used to retain CPS uniformity and avoid
% code duplication.

A \code{Method} definition is a CPS function that takes as input a list \code{[Ref]} of the method's parameters (passed by reference),
the callee object named \code{this}, a \emph{writeback} reference (\verb+Maybe Ref+), and last its current continuation \code{Stm}. 
In case of synchronous call the callee method indirectly writes the \code{Return} value
to the writeback reference of the heap and the execution jumps back to the caller by invoking
the method's continuation; in case of asynchronous call the writeback is empty, the return value is stored
to the caller's future (destiny) and the method's continuation is invoked resulting to the exit of the ABS process.
An object or future reference \code{Ref} is
represented by an integer index to the program's global heap array;
similarly, an object attribute \code{Attr} is an integer index to an internal-to-the-object attribute array, hence
shallow-embedded (compared to embedding the actual name of the attribute). 
Values (\code{V}) in our language can be this-object attributes (A), parameters to the method (P),
integer literals (\code{I}), and integer arithmetic on those values (\code{Add}, \code{Sub}...).
The right-hand side (\code{Rhs}) of an assignment directly reflects that of the source language.
Boolean expressions are only appearing as predicates to \code{If} and \code{While} and are inductively constructed
by the datatype \code{B}, that represents reference and integer comparison.

\begin{figure}[t]
\begin{minipage}{0.35\textwidth}
\begin{lstlisting}[basicstyle=\small\ttfamily]
even x = eq (mod x 2) 0
main = print (even 3) 
\end{lstlisting}
\end{minipage}
\begin{minipage}{0.63\textwidth}
\begin{lstlisting}[numbers=none,basicstyle=\small\ttfamily]
mod' x y k = k (mod x y)
eq' x y k = k (eq x y)
even' x k = mod' x 2 (*'$\lambda$'* a *'$\to$'* eq' a 0 k)
main = even' 3 (*'$\lambda$'* a *'$\to$'* print a)
\end{lstlisting}
\end{minipage}
\secbeg\secbeg\secbeg\secbeg
\caption{(a) Example program in direct style and (b) translated to CPS}
\label{fig:cps}
\end{figure}

% The Haskell language (statically-typed, purely-functional) enabled us 
% to embed (implement) our language in such a way that the end programs
% have reasonable execution speed.
% Moreover, 
% %are amenable to verification and analysis. % , because of the high-level of abstractions.
% thanks to our soundness result (see Sec.~\ref{sec:correctness}) we can transfer the results of ABS analysis tools to the Haskell translated code.

% The complete code as well as some sample compilations can be found at %\url{https://github.com/bezirg/abs2haskell-pure/tree/ast}.
% \url{https://github.com/abstools/abs-haskell-formal}.

\begin{figure}[t]
\begin{lstlisting}[language=hs,numbers=none, basicstyle=\small\ttfamily]
type Method = [Ref] *'$\to$'* Ref *'$\to$'* Maybe Ref *'$\to$'* _Stm_ *'$\to$'* +Stm+
data Stm where -- (formatted in GADT syntax)
   Skip :: _Stm_ *'$\to$'* +Stm+
   Await :: Attr *'$\to$'* _Stm_ *'$\to$'* +Stm+
   Assign :: Attr *'$\to$'* Rhs *'$\to$'* _Stm_ *'$\to$'* +Stm+
   If :: B *'$\to$'* (_Stm_*'$\to$'*+Stm+) *'$\to$'* (_Stm_*'$\to$'*+Stm+) *'$\to$'* _Stm_ *'$\to$'* +Stm+
   While :: B *'$\to$'* (_Stm_*'$\to$'*+Stm+) *'$\to$'* _Stm_ *'$\to$'* +Stm+
   Return :: Attr *'$\to$'* Maybe Ref *'$\to$'* _Stm_ *'$\to$'* +Stm+
\end{lstlisting}
\secbeg\secbeg\secbeg\secbeg
\begin{minipage}{0.50\textwidth}
\begin{lstlisting}[language=hs,numbers=none, basicstyle=\small\ttfamily]
data Rhs = Val V
         | New
         | Get Attr
         | Async Attr Method [Attr]
         | Sync Method [Attr]
\end{lstlisting}
\end{minipage}
\secbeg\secbeg\secbeg
\begin{minipage}{0.62\textwidth}
\begin{lstlisting}[language=hs,numbers=none, basicstyle=\small\ttfamily]
type Ref = Int
type Attr = Int
data B = B *'$:\!\wedge$'* B | B *'$:\!\vee$'*  B | *'$:\!\neg$'* B | V *'$:\equiv$'* V
data V = A Ref | P Ref | I Int
        | Add V V | Sub V V ...
\end{lstlisting}
\end{minipage}
\secbeg\secbeg\secbeg
\caption{The syntax and types of the target language.
Continuations are \uwave{wave-underlined}. The program/process final result type is \uuline{double-underlined}}
\label{fig:stm}
\end{figure}

\begin{figure}[t]
%\surbracket{S}_{s} =&~ \surbracket{S}_{s,\texttt{undefined},\texttt{Nothing}}\\
%\surbracket{S}_{s,k} =&~ \surbracket{S}_{s,k,\texttt{Nothing}}\\
% \trStmt{S}{k,wb} =&~ \left\{
% \begin{array}{ll}
% \trStmt{S_1}{k',wb} & \text{~~~if } S=S_1;S_2, ~~k' = \trStmt{S_2}{k,wb}\\
% \trStmt{S}{k,wb} & \text{~~~i.o.c }\\
% \end{array} \right.\\[1em]
\begin{align*}
\trStmt{\texttt{x:=}V}{k,wb} =&~ \texttt{Assign}~x~\trV{V}~k 	
  & \trStmt{\texttt{skip}}{k,wb} =&~ \texttt{Skip}~k\\[-.1cm]
\trStmt{\texttt{x:=new}}{k,wb} =&~ \texttt{Assign}~x~\texttt{New}~k
  & \trStmt{\texttt{await~f}}{k,wb} =&~ \texttt{Await}~f~k\\[-.1cm]
\trStmt{\texttt{x:=f.get}}{k,wb} =&~ \texttt{Assign}~x~(\texttt{Get}~f)~k
  & \trStmt{\texttt{return~x}}{k,wb} =&~ \texttt{Return}~x~wb~k\\[-.1cm]
\trStmt{\texttt{x:=y!m($\bar{z}$)}}{k,wb} =&~ \texttt{Assign}~x~(\texttt{Async}~y~m~\bar{z})~k 
  & \trStmt{\texttt{return$^*$~x}}{k,wb} =&~ \texttt{Return}~x~\texttt{Nothing}~k\\[-.1cm]
\trStmt{\texttt{x:=m($\bar{z}$)}}{k,wb} =&~ \texttt{Assign}~x~(\texttt{Sync}~m~\bar{z})~k 
  & \trStmt{\texttt{return$^z$~x}}{k,wb} =&~ \texttt{Return}~x~(\texttt{Just}~z)~k\\[-.1cm]
\trStmt{S_1;S_2}{k,wb} =&~ \trStmt{S_1}{k',wb} \text{~with } k' = \trStmt{S_2}{k,wb}\\[-1.1cm]
\end{align*}
\begin{align*}
\trStmt{\mathtt{if}\ B\ \{ S_1 \} \ \mathtt{else}\  \{ S_2 \}}{k,wb} =~ \texttt{If}~\trB{B}~(\backslash k' \to \trStmt{S_1}{k',wb})~(\backslash k' \to \trStmt{S_2}{k',wb})~k\\[-.1cm]
\trStmt{\mathtt{while} \ B\ \{ S \}}{k,wb} = \texttt{While}~\trB{B}~(\backslash k' \to \trStmt{S}{k',wb})~k~~~~~~~~~~~~~~~~~~~~~~~\\\\[-.1cm]
\end{align*}
% \vspace{-1.3cm}
% \begin{align*}
% \trStmt{S_1;S_2}{k,wb} =&~ \trStmt{S_1}{k',wb} \text{~~~where } k' = \trStmt{S_2}{k,wb}\\
% \end{align*}
\vspace{-1.5cm}
\begin{align*}
\trMethod{m} =&~ (\mathtt{m~l~this~wb~k =~} \trStmt{S_m}{\mathtt{k},\mathtt{wb}}) \\[-.1cm]
& \text{where}~m(\bar{w}) \mapsto S_m \in D ~\text{and \texttt{l} is the Haskell list  
 that contains } \\[-.1cm] & \text{the same elements as the sequence}~\bar{w}\\
\end{align*}
\secbeg\secbeg\secbeg\secbeg\secbeg\secbeg
\secbeg\secbeg\secbeg\secbeg\secbeg\secbeg
\caption{Translation of ABS-subset programs to Haskell AST}
\label{fig:comp_instr}
\end{figure}

The compilation of statements is shown in Fig.~\ref{fig:comp_instr}. The translation $\trStmt{S}{k,wb}$ takes two arguments: the continuation $k$ and the writeback reference $wb$. Each statement is translated into its Haskell counterpart, followed
by the continuation $k$. The multiple rules for the \texttt{return} statement are due to the different uses of the translation: when compiling
methods the \texttt{return} statement will appear unmarked, so we include the writeback passed as an argument; otherwise it is used to translate runtime configurations, so \texttt{return} statements will appear marked and we generate the writeback related to the mark. When omitted, we assume the default values $k=\mathtt{undefined}$ and $wb=\mathtt{Nothing}$ for the  $\trStmt{S}{k,wb}$ translation. $\trB{B}$ represents the translation of a boolean expression $B$, and $\trV{V}$ the translation of integer expressions, references or variables. A method definition translates to a Haskell function that includes the compiled body.

\begin{figure}[t]
\begin{minipage}{0.5\textwidth}
\begin{lstlisting}[basicstyle=\small\ttfamily]
main, map, reduce *'${::}$'* Method
main [] this wb k = 
  Assign node1 New *'\textit{\$}'* 
  Assign node2 New *'\textit{\$}'* 
  Assign f1 (Async node1 map [*'$v_1$'*])*'\textit{\$}'* 
  Assign f2 (Async node2 map [*'$v_2$'*])*'\textit{\$}'*
  Await f1 *'\textit{\$}'*  
  Await f2 *'\textit{\$}'* 
  Assign r1 (Get f1) *'\textit{\$}'* 
\end{lstlisting}
\end{minipage}
\begin{minipage}{0.4\textwidth}
\begin{lstlisting}[language=hs,firstnumber=10, basicstyle=\small\ttfamily]
  Assign r2 (Get f2) *'\textit{\$}'* 
  Assign r (Sync reduce [r1,r2]) *'\textit{\$}'*  
  Return r wb k
  
map [v] this wb k = ...
reduce [a,b] this wb k = ...

-- Position in the attribute array
[node1,node2,f1,f2,r1,r2,r] = [0..]
\end{lstlisting}
\end{minipage}
\secbeg\secbeg\secbeg
\caption{The Haskell-translated running example of MapReduce\label{fig:running}}
\label{fig:tar}
\end{figure}

% \begin{definition}[Translation of methods]\label{def:transMethod}
%  Consider a method $m(\bar{w}) \mapsto S_m \in D$ and the Haskell list \texttt{l} 
%  that contains the same elements as the sequence $\bar{w}$. The translation of
%  method $m$ is the Haskell function of type \code{Method}:
%  
%  \texttt{m l this wb k = } $\surbracket{S_m}_{s,\mathtt{k},\mathtt{wb}}$\\
% %  $$
% %  \begin{array}{ll}
% %   \texttt{m}~\bar{r}~\texttt{this}~\texttt{Nothing}~k &=~ \surbracket{S_m\tau}_{s,k}\\
% %   \texttt{m}~\bar{r}~\texttt{this}~(\texttt{Just}~x)~k &=~ \surbracket{\widehat{S_m\tau}^x}_{s,k}\\
% %  \end{array}
% %  $$  
%  \end{definition}
%  
% the original program of Fig.~\ref{fig:src} to the target language
% 
% in Fig.~\ref{fig:tar}.
% %derived from the one-to-one denotational semantics of this transcompilation given in Fig.~\ref{fig:comp_instr}.
% This compilation uses the translation of statements in Fig.~\ref{fig:comp_instr} and the following
% definition:

%\secbeg\secbeg\secbeg\secbeg
 \subsection{Runtime execution}\label{sec:runtime} 
\secbeg
The program heap is implemented as the triple: array of objects, array of futures and a \verb+Int+ counter.
Every cell in the objects-array designates 1 object holding a pair of its attribute array and process queue (double-ended)
in Haskell \verb+IOVector (IOVector Ref,+ \verb+Seq Proc)+.
A cell in futures-array denotes a future which is either unresolved with a number of listener-objects \verb+await+ing for it to be completed,
or resolved with a final value, i.e. \verb+IOVector (Either [Ref] Ref)+. An ever-increasing counter
is used to pick new references; when it reaches the arrays' current size both of the arrays double in size (i.e. dynamic arrays).
The size of all attribute arrays, however, is fixed and predetermined at compile-time, 
by inspecting the source code (as shown in L18 of Fig.\ref{fig:tar}).
%; we assume a prior static analysis that maps attribute names to indexes for the attribute array.

An \verb+eval+ function accepts a \code{this} object reference and the current heap and executes
a single statement of the head process in the process queue, returning a new heap and those objects that have become active after the execution
(\verb+eval this heap :: IO+ \verb+(Heap, [Ref]+).
%an active object has at least one process in the process queue and is not blocked on a ``future.get'' call. 
An \verb+await+ executed statement will put its continuation (current process) in the tail
of the process queue, effectively enabling cooperative multitasking, whereas all others will keep it as the head. A \verb+Return+
executed statement originating from an asynchronous call is responsible for re-activating the objects that are blocked on its resolved future.
A global scheduler ``trampolines'' over a queue of active objects: it calls \verb+eval+ on the head object, puts the newly-activated objects in the tail of the queue, and loops until no objects are left in the queue---meaning the ABS program is either finished or deadlocked.
At any point in time, the pair of the scheduler's object queue with the heap comprise the program's state. 

\paragraph{Comparison.}
The described target language is an untyped extract of the canonical ABS-Haskell backend~\cite{bezirg_cloud_2016},
with the main difference being that ABS statements are translated to an AST interpreted by \verb+eval+ function,
while the canonical version compiles statements down to native code, which naturally yields faster execution.
However, this deep embedding of an AST allows multiple interpretations of the syntax: debug the syntax tree and have an equivalence result.
At runtime, the \code{eval} function operates in ``lockstep'' (i.e. executing one CPS statement at a time)
whereas the canonical backend applies CPS between release points
%only from release point to release point
(\code{await}, \code{get} and \verb+return+ from asynchronous calls) which benefits in performance
but would otherwise make reasoning about correctness and resource preservation for this setup more involved.
Another argument for lockstep execution is that we can ``simulate'' a global Haskell-runtime scheduler (with a N:1 threading model)
and include it in our proofs, instead of reasoning for the lower-level C internals of the GHC runtime thread scheduler (with M:N parallelism).

% ENRIQUE: added for LOPSTR'16
Our target language is also related to \emph{Coroutining Logic Engines} presented in~\cite{Tarau2011} for concurrent Prolog. These engines %are high-level constructs that
encapsulate multi-threading by providing entities that evaluate goals and yield answers when requested. They follow a similar coroutining approach, % to asynchronous methods,
however, logic engines can produce several results, whereas asynchronous methods can be suspended by the scheduler many times but they only generate one result when they finish.

\section{Correctness and Resource Preservation}\label{sec:correctness}
\secbeg\secbeg
% \todo{EM: This section could be reduced a bit (moving some results to the appendix) or a lot (moving even the intermediate semantics and re-stating the theorems)}

\begin{figure}[!t]
{\footnotesize
$$
% \inference[(\textsc{Assign I})]
% {
% \nextO{h}{[\overline{o_m}]} = o_n & h(o_n)(\procq) = (\texttt{Assign}~x~(\texttt{Attr}~y)~k',l)\cdot q \\
% %k(\emptyset) = \texttt{Assign}~x~(\texttt{Attr}~y)~k' \\
% h' = h[ (o_n)(x) \mapsto h(o_n)(y), (o_n)(\procq) \mapsto (k',l)\cdot q]\\
% }
% {
% (h,[\overline{o_m}]) \toH (h',[\overline{o_{n+1\to m}}]:[\overline{o_{1\to n}}])
% }
% $$
\inference[(\textsc{Assign})]
{
\nextO{h}{[\overline{o_m}]} = o_n & h(o_n)(\procq) = (\texttt{Assign}~x~V~k',l)\cdot q \\
\getVal{h(o_n)}{V} = v & h' = h[ (o_n)(x) \mapsto v, (o_n)(\procq) \mapsto (k',l)\cdot q]\\
}
{
(h,[\overline{o_m}]) \toH (h',[\overline{o_{n+1\to m}}]:[\overline{o_{1\to n}}])
}
$$
%
% $$
% \inference[(\textsc{Assign II})]
% {
% \nextO{h}{[\overline{o_m}]} = o_n & h(o_n)(\procq) = (\texttt{Assign}~x~(\texttt{Param}~r)~k',l)\cdot q \\
% %k(\emptyset) = \texttt{Assign}~x~(\texttt{Param}~r)~k' \\
% h' = h[ (o_n)(x) \mapsto r, (o_n)(\procq) \mapsto (k',l)\cdot q]\\
% }
% {
% (h,[\overline{o_m}]) \toH (h',[\overline{o_{n+1\to m}}]:[\overline{o_{1\to n}}])
% }
% $$
%
$$
\inference[(\textsc{New})]
{
\nextO{h}{[\overline{o_m}]} = o_n & h(o_n)(\procq) = (\texttt{Assign}~x~\texttt{New}~k',l)\cdot q \\
%k(\emptyset) = \texttt{Assign}~x~\texttt{New}~k' & 
h(count)= o_{new} &
% Sched' = \texttt{fromList}~\redt{(o_m,[~])}:[\ldots , (o_n, (k',l):q), \ldots] \\
%Sched' = \texttt{fromList}~[\ldots , (o_n, (k',l):q), \ldots] \\
h' = h[ (o_n)(x) \mapsto o_{new}, count \mapsto o_{new}+1, \\%(o_{new}) \mapsto \epsilon, \\ 
~~~~~~~~~~~~~~~~(o_{new})(\procq) \mapsto \epsilon, (o_n)(\procq) \mapsto (k',l)\cdot q]}
{
(h,[\overline{o_m}]) \toH (h',[\overline{o_{n+1\to m}}]:[\overline{o_{1\to n}}])
%\langle \texttt{fromList}~[\ldots , (o_n, (k,l):q), \ldots], h \rangle \rightarrow \langle Sched', h' \rangle
}
$$
$$
\inference[(\textsc{Get})]
{
\nextO{h}{[\overline{o_m}]} = o_n & h(o_n)(\procq) = (\texttt{Assign}~x~(\texttt{Get}~f)~k',l)\cdot q \\
%k(\emptyset) = \texttt{Assign}~x~(\texttt{Get}~f)~k' & h(h(o_n)(f)) = \texttt{Just}~v \\
%k(\emptyset) = \texttt{Assign}~x~(\texttt{Get}~f)~k' & 
h(h(o_n)(f)) = \texttt{Right}~v &
h' = h[ (o_n)(x) \mapsto v, (o_n)(\procq) \mapsto (k',l)\cdot q]
}
{
(h,[\overline{o_m}]) \toH (h',[\overline{o_{n+1\to m}}]:[\overline{o_{1\to n}}])
}
$$
$$
\inference[(\textsc{Await I})]
{
\nextO{h}{[\overline{o_m}]} = o_n & h(o_n)(\procq) = (\texttt{Await}~f~k',l)\cdot q \\
% k(\emptyset) = \texttt{Await}~f~k' & h(h(o_n)(f)) = \texttt{Just}~v \\
%k(\emptyset) = \texttt{Await}~f~k' & 
h(h(o_n)(f)) = \texttt{Right}~v &
h' = h[(o_n)(\procq) \mapsto (k',l)\cdot q]\\
}
{
(h,[\overline{o_m}]) \toH (h',[\overline{o_{n+1\to m}}]:[\overline{o_{1\to n}}])
}
$$
$$
\inference[(\textsc{Await II})]
{
\nextO{h}{[\overline{o_m}]} = o_n & h(o_n)(\procq) = (\texttt{Await}~f~k',l)\cdot q \\
% k(\emptyset) = \texttt{Await}~f~k' & h(h(o_n)(f)) = \texttt{Nothing} \\
%k(\emptyset) = \texttt{Await}~f~k' & 
h(h(o_n)(f)) = \texttt{Left}~e & 
h' = h[(o_n)(\procq) \mapsto q\cdot (\texttt{Await}~f~k',l)]\\
%Sched' = \texttt{fromList}~[\ldots , (o_n, q:(\lambda~\emptyset \to \texttt{Await}~f~k',l)), \ldots] \\
}
{
(h,[\overline{o_m}]) \toH (h',[\overline{o_{n+1\to m}}]:[\overline{o_{1\to n}}])
}
$$
$$
\inference[(\textsc{Async})]
{
\nextO{h}{[\overline{o_m}]} = o_n & h(o_n)(\procq) = (\texttt{Assign}~f~(\texttt{Async}~x~m~\bar{z})~k',l)\cdot q \\
h(count) = l' &
%k(\emptyset) = \texttt{Assign}~x~(\texttt{Async}~y~m~\bar{z})~k' & 
h(o_n)(x) = o_x & h(o_x)(\procq) = q_x & (m(\bar{w}) \mapsto S) \in D\\
k'' = \texttt{m}~h(o_n)(\bar{z})~o_n~\texttt{Nothing}~\texttt{undefined} &
%\tau = [\bar{w} \mapsto h(o_n)(\bar{z})] & \\
%Sched' = \texttt{fromList}~[\ldots , (o_n, (k',l):q), \ldots,(o_m, q_m:(k'',l')),\ldots] \\
\newQadd([\overline{o_m}],o_n,o_x) = s\\
%  h' = h[ (o_n)(x) \mapsto l', count \mapsto l'+1, l' \mapsto \bot, \\
h' = h[ (o_n)(f) \mapsto l', count \mapsto l'+1, l' \mapsto \texttt{Left}~[~], \\
 (o_n)(\procq) \mapsto (k',l)\cdot q, (o_x)(\procq) \mapsto q_x\cdot (k'',l')]
}
{
(h,[\overline{o_m}]) \toH (h',s)
% \langle \texttt{fromList}~[\ldots , (o_n, (k,l):q), \ldots,(o_m, q_m),\ldots], h \rangle \rightarrow \langle Sched', h' \rangle
}
$$
%\redt{Similar if $o_n = o_m$, but appending the task into the same queue $q$}
%
$$
\inference[(\textsc{Sync})]
{
\nextO{h}{[\overline{o_m}]} = o_n & h(o_n)(\procq) = (\texttt{Assign}~x~(\texttt{Sync}~m~\bar{z})~k',l)\cdot q \\
%k(\emptyset) = \texttt{Assign}~x~(\texttt{Sync}~m~\bar{z})~k' & 
(m(\bar{w}) \mapsto S) \in D & 
k'' = \texttt{m}~h(o_n)(\bar{z})~o_n~(\texttt{Just}~x)~k' &
h' = h[(o_n)(\procq) \mapsto (k'',l)\cdot q]
}
{
(h,[\overline{o_m}]) \toH (h',[\overline{o_{n+1\to m}}]:[\overline{o_{1\to n}}])
%\langle \texttt{fromList}~[\ldots , (o_n, (k,l):q), \ldots], h \rangle \rightarrow \langle Sched', h \rangle
}
$$
$$
\inference[(\textsc{Return$_A$})]
{
\nextO{h}{[\overline{o_m}]} = o_n & h(o_n)(\procq) = (\texttt{Return}~x~\texttt{Nothing}~\_,l)\cdot q \\
%k(\emptyset) = \texttt{Return}~z~\texttt{Nothing}~\_ & 
\newQdel([\overline{o_m}],o_n,q) = s &
h' = h[l \mapsto \texttt{Right}~h(o_n)(x), (o_n)(\procq) \mapsto q] \\
% ~\\
% Sched' = \texttt{fromList}~[\ldots , (o_n, q), \ldots] & h' = h[l \mapsto h(o_n)(z)]\\
}
{
(h,[\overline{o_m}]) \toH (h',s)
}
$$
%
% \redt{We may need an additional rule when returning from the last task, to remove the entry of $o_n$ from $Sched'$}
%
$$
\inference[(\textsc{Return$_S$})]
{
\nextO{h}{[\overline{o_m}]} = o_n & h(o_n)(\procq) = (\texttt{Return}~x~(\texttt{Just~z})~k',l)\cdot q \\
%k(\emptyset) = \texttt{Return}~z~(\texttt{Just~x})~k'\\
 h' = h[(o_n)(z) \mapsto h(o_n)(x),
(o_n)(\procq) \mapsto (k',l)\cdot q]\\
}
{
(h,[\overline{o_m}]) \toH (h',[\overline{o_{n+1\to m}}]:[\overline{o_{1\to n}}])
}
$$
}
\secbeg\secbeg\secbeg
\caption{Intermediate semantics.}
\label{fig:highlevelHaskell}
\end{figure}

To prove that the translation is correct and resource preserving, we use an intermediate semantics
$\toH$ closer to the Haskell programs. This semantics, depicted in Fig.~\ref{fig:highlevelHaskell},
considers configurations $(h,[\overline{o_m}])$ where all the information of the objects 
is stored in a unified heap---concretely $h(o_n)(\procq)$ returns the process queue 
of object $o_n$. The semantics in Fig.~\ref{fig:highlevelHaskell} presents two
main differences w.r.t. that in Fig.~\ref{fig:localSem} and~\ref{fig:globalSem} of Sec.~\ref{sec:lang}. First, the list $[\overline{o_m}]$ is used
to apply a \emph{round-robin} policy: the first unblocked object\footnote{Object whose
%An object is unblocked if its 
active process is not waiting for a future variable in a \code{get} statement.} 
$o_n$ in $[\overline{o_m}]$
is selected using $\nextO{h}{\overline{[o_m]}}$, the first statement of the active process of $o_n$ is executed and then
the list is updated to continue with the object $o_{n+1}$. The other difference
is that process queues do not contain sequences of statements but 
\emph{continuations}, as explained in the previous section. To generate these
continuation rules (\textsc{Async}) and (\textsc{Sync}) invoke the translation 
of the methods \code{m} with the adequate parameters.
%\todo{EM: revise that continuations are actually explained} 
Nevertheless, the rules of the 
$\toH$ semantics correspond with the semantic rules in %of the $\to$ semantics in 
Sec.~\ref{sec:lang}.  

Given a list $[\overline{o_m}]$ we use the notation $[\overline{o_{i\to k}}]$ for the sublist $[o_i,o_{i+i},\ldots,o_k]$, and the operator $(:)$ for list concatenation.
In the rules \textsc{(Async)}
and \textsc{(Return$_A$)}, where the object list can increase or decrease one
object, we use the following auxiliary functions. $\newQadd([\overline{o_m}],o_n,o_y)$ inserts the
object $o_y$ into $[\overline{o_m}]$ if it is new (i.e., it does not appear in $[\overline{o_m}]$), and $\newQdel([\overline{o_m}],o_n,q_n)$ removes the object $o_n$ from $[\overline{o_m}]$ if its process queue $q_n$ is empty. In both cases they
advance the list of objects to $o_{n+1}$.

\secbeg\secbeg
\begin{equation*}
\newQadd([\overline{o_m}],o_n,o_y) = \left\{
\begin{array}{rl}
[\overline{o_{n+1\to m}}]:[\overline{o_{1\to n}}] & \text{~~~if } o_y \in [\overline{o_m}] \\
~[\overline{o_{n+1\to m}}]:[\overline{o_{1\to n}}]:[o_y] & \text{~~~if } o_y \notin [\overline{o_m}]\\
\end{array} \right.
\end{equation*}

\begin{equation*}
\newQdel([\overline{o_m}],o_n,q_n) = \left\{
\begin{array}{rl}
[\overline{o_{n+1\to m}}]:[\overline{o_{1\to n-1}}] & \text{~~~if } q_n = \epsilon \\
~[\overline{o_{n+1\to m}}]:[\overline{o_{1\to n}}] & \text{~~~if } q_n \neq \epsilon \\
\end{array} \right.
\end{equation*}

In order to reason about the different semantics, we define the translation from
runtime configurations $\langle C, h \rangle$ of Sec.~\ref{sec:lang} 
to concrete Haskell data structures used in the intermediate 
$\toH$ semantics and in the compiled Haskell programs (see Fig.~\ref{fig:conf}). The set
of closure lists $C$ is translated into a list of object references, and the process queues 
inside $C$ are included into the heap related to the special term $\procq$.
Although we use the same notation $h$, we consider that the heap is translated into 
the corresponding Haskell tuple (\textit{object\_vector}, \textit{future\_vector}, \textit{counter}) explained in Sec.~\ref{sec:targetlang}.
%where $\mi{ov}$ is a vector of object
%information, $\mi{fv}$ a vector of futures and $c$ a reference counter. 
%
As usual with
heaps, we use the notation $h[(o_n)(\procq) \mapsto q]$ to update the 
process queue of the object $o_n$ to $q$.
%Pairs and triples in the source configuration are translated to Haskell tuples; mathematical functions are
%translated to fast, size-balanced binary trees \verb+(fromList)+; the closure's body $S\tau$ is translated
%to its statements $S$ substituted with the local variable bindings ($\surbracket{S}\backslash\tau$); 
Finally, natural numbers
become integers, global variables become Strings and $\mi{Nat}_\bot$ values in the 
futures become $\mi{Either}$ values.
To denote the inverse translation from data structures to runtime configurations we use 
$\trConfInv{(h',\mi{act})} = \langle C,h \rangle$---the same for queues
$\trQueueInv{\cdot}$ and statements $\trStmtInv{\cdot}$. % in Fig.~\ref{fig:comp_instr}.
Note that the translation $\trConf{\cdot}_c$ is not
deterministic because it generates a list of object references from a set of 
closures $C$, so the order of the objects in the list is not defined. On the other hand, 
the translation of the heap in $\trConf{\cdot}$ and the inverse translation 
$\trConfInv{\cdot}$ are deterministic.

% In order to prove the soundness and completeness of the translation w.r.t. the 
% evaluation we require that the translation of a method is sound, i.e, the 
% related closure actually represents its body.

\begin{figure}[tb]
$\begin{array}{l@{~~~~~~~~~~~~~}ll}
\trConf{\langle C,h \rangle} = ~(h',\mi{act}), \textnormal{where} & \trQueue{\epsilon} &=~ \epsilon \\
  ~~~ \mi{act} = [ o_n ~|~ (o_n,Q_n) \in C, Q_n \neq \epsilon] & \trQueue{(S,l)\cdot Q} &=~ (\trStmt{S},l)\cdot \trQueue{Q} \\
  ~~~ C = \{ (n_1,Q_1), \ldots, (n_m,Q_m) \}~\textnormal{and}\\
  ~~~ h' = h[ \overline{(n_i)(\procq) \mapsto \trQueue{Q_i} } ] \\
\end{array}$
% \begin{array}{lcl}
% \surbracket{\epsilon}_q &=&~ \epsilon \\
% \surbracket{(S,l)\cdot Q}_q &=&~ (\surbracket{S}_s,l)\cdot \surbracket{Q}_q \\
% \end{array}$
\vspace*{-.25cm}
\caption{Translation from source to target configurations.\label{fig:conf}}
\end{figure}
Based on the previous definitions we can state the
soundness of the traces, i.e., every trace of \code{eval} steps is a valid trace w.r.t. $\to$. 
Note that for the sake of conciseness we unify the statements $S$ and their representation as Haskell terms \code{res}, since there is a straightforward translation between them. We consider the auxiliary function
$\updL([\overline{o_m}], o_n, l) = [\overline{o_{n+1 \to m}}]:[\overline{o_{1 \to n-1}}]:l$
to update the list of object references.

 \begin{theorem}[Trace soundness]\label{teo:traceSoundness}
  Let $(h_1,s_1)$ be an initial state and consider a sequence of $n-1$ consecutive \code{eval} steps defined as: 
  a) $o_i = \nextO{h_i}{s_i}$, 
  b) \code{eval o$_i$} \code{h$_i$} = \code{(res$_i$, l$_i$, h$_{i+1}$)}, 
  c) $s_{i+1} = \updL(s_i, o_i, l_i)$. Then 
  $\trConfInv{(h_1,s_1)} \todec{o_1}{\mi{res}_1} \trConfInv{(h_2,s_2)}_c \todec{o_2}{\mi{res}_2} \ldots \todec{o_{n-1}}{\mi{res}_{n-1}} \trConfInv{(h_n,s_n)}$.
  %If $t_1 \toHdec{o_1}{S_1} t_2 \toHdec{o_2}{S_2} \ldots \todec{o_n}{S_n} t_n$ then $\surbracketinv{t_1}_c \todec{o_1}{S_1} \surbracketinv{t_2}_c \todec{o_2}{S_2} \ldots \todec{o_n}{S_n} \surbracketinv{t_n}_c$
 \end{theorem}

Note that it is not possible to obtain a similar result about trace completeness since the $\to$-semantics in Fig.~\ref{fig:globalSem}
selects the next object to execute nondeterministic (random scheduler), whereas the intermediate $\toH$-semantics in Fig.~\ref{fig:highlevelHaskell} follows a concrete \emph{round-robin} scheduling policy.
% (\emph{It's not true because of the round-robin scheduler}) If $A_1 \todec{o_1}{S_1} A_2 \todec{o_2}{S_2} \ldots \todec{o_n}{S_n} A_n$ then there exist $\overline{t_n}$
%   such that $t_i = \surbracket{A_i}_c$ and $t_1 \toHdec{o_1}{S_1} t_2 \toHdec{o_2}{S_2} \ldots \todec{o_n}{S_n} t_n$.
%The proofs of the theorems is include include in Appendix~\ref{sec:proofs}. 
As a final remark notice that the intermediate semantics $\toH$
%, since it is expressed in terms of Haskell constructions, 
can be seen as a \emph{specification} of the \code{eval} function. % for any translated program.
Therefore it can be used to guide the correctness proof of \code{eval} using %semi-automatic 
proof assistance tools like \emph{Isabelle}~\cite{Nipkow2002} or to generate tests automatically using \emph{QuickCheck}~\cite{Claessen2000}.

\subsection{Preservation of Resource Consumption}\label{sec:preservation}
\secbeg
A strong feature of our translation is that the Haskell-translated program preserves
the \emph{resource consumption} of the original ABS program. 
As in~\cite{AlbertACGGPR15} we use the notion of \emph{cost model} to
parameterize the type of resource we want to bound. Cost models are
functions from ABS statements to real numbers, i.e., $\costmodel : S
\to \mathbb{R}$ that define different resource consumption
measures. For instance, if the resource to measure is the number of
executed steps, $\costmodel : S \to {1}$ such that each instruction
has cost one. However, if one wants to measure memory consumption, we
have that $\costmodel(new)=\mathtt{c}$, where $\mathtt{c}$ refers to the
size of an object 
%of type $C$, 
reference,
and $\costmodel(instr)=0$ for all
remaining instructions.
The resource preservation is based on the notion of
\emph{trace cost}, i.e., the sum of the cost of the statements executed.
Given a concrete cost model $\costmodel$, an object reference $o$ and a program execution 
%$\trace \equiv A_1 \todec{o_1}{S_1} A_2 \todec{o_2}{S_2} \ldots \todec{o_{n-1}}{S_{n-1}} A_n$, 
$\trace \equiv A_1 \todec{o_1}{S_1} \ldots \todec{o_{n-1}}{S_{n-1}} A_n$, 
the cost of the trace $\cost(\trace,o,\costmodel)$ is defined as:

\secbeg\secbeg
$$
\cost(\trace,o,\costmodel) = \sum_{S\in\trace|_{\{o\}}} \costmodel(S)
$$

%\secbeg
Notice that, from all the steps in the trace $\trace$, it takes into
account only those performed in object $o$ (denoted as $\trace|_{\{o\}}$), so the cost notion is 
\emph{object-sensitive}. 
%
%The cost of a trace depends only on the instructions executed in the
%objects. 
Since the trace soundness states that the \code{eval} function performs the same steps as some trace $\trace$, the cost preservation is a straightforward corollary:

\begin{corollary}[Consumption Preservation]\label{teo:sameCost}
 Let $(h_1,s_1)$ be an initial state and consider a sequence $\trace_E$ of $n-1$ consecutive \code{eval} steps defined as: 
 a) $o_i = \nextO{h_i}{s_i}$, 
 b) \code{(res$_i$, l$_i$, h$_{i+1}$) = eval o$_i$} \code{ h$_i$}, 
 c) $s_{i+1} = \updL(s_i, o_i, l_i)$. Then 
 %there is a trace 
 %$\trace = \globalTr{(h_1,s_1)} \toC^* \globalTr{(h_n,s_n)}$ such that $\cost(\trace_E,o,\costmodel) = \cost(\trace_C,o,\costmodel)$.
 $\trace = \trConfInv{(h_1,s_1)} \todec{o_1}{\mi{res}_1}
 \trConfInv{(h_2,s_2)}_c \todec{o_2}{\mi{res}_2} \ldots
 \todec{o_{n-1}}{\mi{res}_{n-1}} \trConfInv{(h_n,s_n)}$ such that
 $\cost(\trace_E,o,\costmodel) = \cost(\trace,o,\costmodel)$.
\end{corollary}

% OLD theorem regarding the semantics in [1]
% \begin{theorem}[Consumption Preservation]\label{teo:sameCost}
%  Let $(h_1,s_1)$ be an initial state and consider a sequence $\trace_E$ of $n-1$ consecutive \code{eval} steps defined as: 
%  a) $o_i = \nextO{h_i}{s_i}$, 
%  b) \code{(res$_i$, l$_i$, h$_{i+1}$) = eval o$_i$} \code{ h$_i$}, 
%  c) $s_{i+1} = \updL(s_i, o_i, l_i)$. Then 
%  there is a trace 
%  $\trace_C = \globalTr{(h_1,s_1)} \toC^* \globalTr{(h_n,s_n)}$ such that $\cost(\trace_E,o,\costmodel) = \cost(\trace_C,o,\costmodel)$.
% \end{theorem}
%
As a side effect of the previous result, we know that the upper bounds
that are inferred from the ABS programs (using resource analyzers like~\cite{AlbertACGGPR15}) are
valid upper bounds for the Haskell translated code. We denote by
$\mi{UB_{main}}()|_{o}$ the upper bound obtained for the analysis of
a  \texttt{main} method for the computation performed on object \texttt{o}.

% .  Combining
% Theorem~3 from \cite{AlbertACGGPR15} and Theorem~\ref{teo:sameCost} we
% obtain the soundness of the resource analysis for the Haskell
% translation as follows.

\begin{theorem}[Bound preservation]\label{teo:boundPreservation}
Let $P$ be a program, $\trace_E$ a sequence of \code{eval} steps from an initial state $(h_1,s_1)$ and $\mi{UB_{main}}()|_{o}$ the 
upper bound obtained for the program $P$ starting from the main block, restricted to the object $o$. Then
 %If~$\trace_H = \mi{A_1} \toHdec{o_1}{S_1} \mi{A_2} \toHdec{o_2}{S_2} \ldots \toHdec{o_{n-1}}{S_{n-1}} \mi{A_n}$ then 
 $\cost(\trace_E,o,\costmodel) \leq \mi{UB_{main}}()|_{o}$
\end{theorem}

% \subsection{Preservation of execution steps}
% \todo{EM: explain that if the Haskell implementation performs a constant number of $\beta$-reductions---$O(1)$---to execute one ABS statement, then the 
% number of $\beta$-reductions executed by object $o$ in any execution starting from \emph{main} is in $O(\mi{UB_{main}}()|_{\mi{name(o)}})$}

%\FloatBarrier

%%% Local Variables:
%%% mode: latex
%%% TeX-master: "0_main"
%%% End:

\section{Experimental Evaluation}\label{sec:experiments}
\secbeg\secbeg
\begin{figure}[tb]
 \includegraphics[scale=0.96]{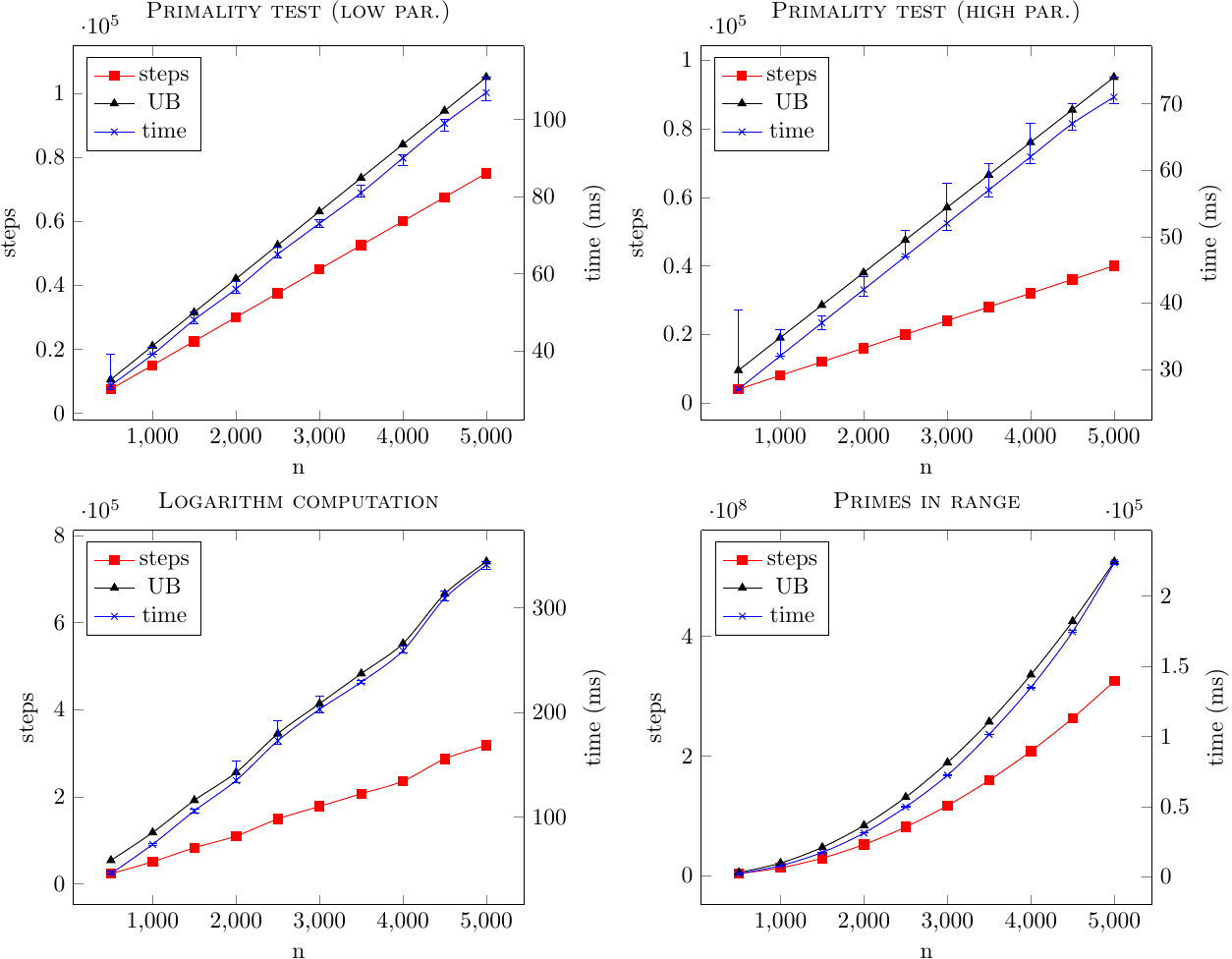}
 %\input{plots}
 %\vspace*{-.75cm}
 \caption{Execution steps vs. time ({\small Intel\textsuperscript{\textregistered}~Core\textsuperscript{TM} i7-4790 at 3.60GHz, 16 GB}).\label{fig:plots}}
\end{figure}
% It might be a good idea to separate steps-UB and time in different plots 
% because with error bars it will too crowded => linear & quadratic only

In the previous section we proved that the execution of compiled Haskell
\cbstart programs has the same resource consumption as the original ABS traces wrt.
any concrete cost model $\costmodel$, i.e., 
both programs execute the same ABS statements in the same order and in the same objects. 
However, cost models are defined in terms of ABS statements so they are unaware
of low-level details of the Haskell runtime environment as $\beta$-reductions
or garbage collection.
% For example, the execution of one ABS instruction can require several $\beta$-reductions
Studying the relation between cost models and some significant low-level details of 
the Haskell runtime
in a formal way is an interesting line of future work. In this section we 
address empirically one particular topic: the Haskell runtime does not introduce 
additional overhead, i.e., the execution of one ABS statement requires only a 
constant amount of work. \cbend
In
order to evaluate this hypothesis, we have elaborated programs\footnote{The
ABS-subset experimental programs and measurements together with the 
target language \& runtime reside at \url{http://github.com/abstools/abs-haskell-formal}.}
with different asymptotic costs and measured the number of statements executed
(steps) and their run-time. 
%Fig~\ref{fig:plots} contains all the details.
%
The \emph{Primality test} computes the primality of a number $n$: \cbstart the program creates 
$n$ objects and checks \cbend every possible divisor of $n$ on each object. The 
difference is that the \emph{low paralellism} version awaits for the result of 
one divisor before invoking the next check and the \emph{high parallelism} 
version does not. Both programs have a $O(n)$ cost.
The \emph{Logarithm computation} program computes the integer part $n$
logarithms. It has cost $O(n.\mi{log}~n)$.
Finally \emph{Primes in a range} computes the prime numbers in the interval 
$[1..n]$, thus having a $O(n^2)$ cost.
%
%programs performs different computations: create a number $n$ of objects (size) and invoke 
%some tasks in each one: 1 task for the linear programs, $\mi{log}~n$ tasks for the linearithmic
%program and $n$ tasks for the quadratic program.
%The difference between the two linear programs is that the \emph{low parallelism}
%version awaits for the result of the task before creating the next object, whereas
%the \emph{high parallelism} version does not await.

We have tested the programs with $n$ ranging from $500$ to $5000$, running $20$
experiments for every value of $n$, and measured the time.
% the mode, minimum and maximum time. 
This is plotted in the cross line (right margin) in 
Fig.~\ref{fig:plots}. The plot
represents the mode times and the minimum and maximum times as
\emph{whiskers}. We have also measured the actual number of steps, represented
in the square line (left margin) in Fig.~\ref{fig:plots}. 
These two plots show that the execution time
and the number of executed steps grows with a similar rate in all the programs,
independently of their asymptotical cost, thus confirming that 
%the execution of one statement requires a constant amount of time. % and 
the compilation does not incur any overhead.
%
% Fig.~\ref{fig:plots} shows the results of the tests as graphs, where the left vertical axis
% is used for the number of steps and the right vertical axis for the run-time in 
% milliseconds. The graphs show that both the
% steps and time plots have the same growth rate in all the programs thus confirming
% the proportionality, i.e., the execution of one statement requires a constant 
% amount of time. 

We have also plotted the resource bounds obtained by the SACO 
tool~\cite{SACO} for the different values of $n$ (triangle line, left margin in
Fig.~\ref{fig:plots}). 
\cbstart
SACO can analyze full ABS programs and thus also the subset considered in this
paper, and allows the selection of the cost model of interest.
\cbend
In this case we have analyzed 
the original ABS programs using the cost model that obtains the number of 
\cbstart ABS statements executed. \cbend
As can be appreciated, the upper bounds are sound and overapproximate the actual
number of \cbstart executed statements. The difference between the upper bounds 
and the actual number of statements executed \cbend
is explained for two reasons. First, the SACO tool considers constructor
methods, i.e., methods that are invoked on every new object, %but the subset of 
%the ABS language presented in this paper does not. Therefore, 
so the SACO tool will
count a constant number of extra \cbstart statements \cbend whenever a new object is created.
%corresponding to the invocation and execution of the implicit constructor. 
However, the main source of imprecision are branching points where 
SACO combines different fragments of information.
%, and sometimes it make some simplifications.
%and therefore generating bigger upper bounds. 
A clear example are loops like the one in the \emph{Primes in a range} program. 
The main loop checks if a number $i \in [1..n]$ is a prime number on each iteration, 
and this check needs \cbstart the execution of $i$ statements. \cbend In this situation SACO considers that  
every iteration has the maximum cost ($n$ statements) and generate an upper bound of 
$n^2$ instead of the more precise (but asymptotically equivalent) expression 
$1+2+\ldots +n$.
\secbeg\secbeg
\section{Conclusion and Future Work}\label{sect:conclusion}
\secbeg\secbeg
We have presented a concurrent object-oriented language (a subset of ABS) and its
compilation to Haskell using continuations.
The compilation is formalised in order to establish that the program
behaviour and the resource consumption are preserved
by the translation.
Compared to the only other formalised ABS backend~\cite{JohnsenHSSS10} (in Maude), our Haskell
translation admits the preservation of resource consumption, and
as a side benefit, makes uses of an overall faster backend.\footnote{\url{http://abstools.github.io/abs-bench} keeps an up-to-date benchmark of all ABS backends.}

In the future we plan to extend our formalisations to accommodate full ABS,
both in terms of the omitted parts of the language as well as the 
non-deterministic behaviour of a multi-threaded scheduler, e.g. by broadening our simulated scheduler to 
non-determinism, and perhaps (M:N) thread parallelism.
%Another consideration is to accommodate our resource-preservation result to 
Another consideration is to relate our resource-preservation result to 
a distributed-object extension of ABS~\cite{bezirg_cloud_2016};
specifically, how the resource analysis translates
to network transport costs after any network optimizations or protocol limitations.
Finally, we plan to formally relate the ABS cost 
models used to define the cost of a trace and some of the low-level runtime 
details of the Haskell runtime like $\beta$-reductions, garbage collections
or main memory usage. Thus, we could express trace costs and upper bounds in 
terms closer to the actual running environment.

% We achieved this through a straightforward (one-to-one) mapping of source to target configurations,
% and statements to Haskell expressions. 
% %
% We show that this compilation does not introduce any overhead during execution, so that run-times are proportional to the number of steps executed, and that it outperforms previous compilers of ABS when
% considering the full set of the language.
% %
% The translation is presented only for the core subset of the ABS
% language; it lacks features such as Algebraic Datatypes, ``true''
% non-determinism, and multicore.  
% Normally, to implement true non-determinism and multicore one needs system-level
% threads.  This is indeed the case with our full-blown ABS-to-Haskell
% compiler.

%%% Local Variables:
%%% mode: latex
%%% TeX-master: "0_main"
%%% End:

%\input{highlevel_semantics}

%\bibliographystyle{splncs03}
%\bibliography{refs}

%\newpage
\appendix

%\fbox{\parbox[b]{.97\linewidth}{
%{\em {\bf \em Note for the reviewers}: The following appendix contains the complete proofs of the theoretical results. It is not part of the paper. In case the paper is accepted, it will be made available as a technical report.}}}

%\input{proofs}
%\section{Proofs and auxiliary results}\label{sec:proofs}
%The complete proofs of the theoretical results can be found in the extended version of %this paper, which 
%can be found in %\url{http://gpd.sip.ucm.es/enrique/publications/actors_haskell_extended.pdf}

\end{document}